\documentclass[10pt]{article}
\input epsf

\newcommand{\AaA}   {Astron. \& Astrophys.}
\newcommand{\ApJ}   {Astrophys.~J.}
\newcommand{\AZh}   {Astron. Zhurn.}

\begin{document}

\hspace{1.0cm} \parbox{15.0cm}{

\baselineskip = 15pt

\noindent {\bf THERMAL X-RAY EMISSION OF THE REMNANTS  \\
OF ASPHERICAL SUPERNOVA EXPLOSIONS}

\bigskip
\bigskip

\noindent {\bf O.~Petruk$^1$}

\bigskip

\baselineskip = 9.5pt

\noindent {\small \copyright~2000}

\smallskip

$^1$\noindent {\small {\it 
Institute for Applied Problems in Mechanics and Mathematics \\
National Academy of Sciences of Ukraine \\
3-b Naukova St., 79053 Lviv, Ukraine}} \\
\noindent {\small {\it e-mail:}} {\tt petruk@astro.franko.lviv.ua}

\baselineskip = 9.5pt \medskip

\medskip \hrule \medskip

\noindent Evolution of adiabatic remnants of an aspherical supernova 
explosion in uniform medium are considered. 
Thermal X-ray emission of such remnants 
are investigated. It is shown that integral thermal X-ray 
characteristics (X-ray luminosity and spectrum) 
of the objects do not allow us to reveal the assymetry in the explosion 
because these characteristics are close to their Sedov counterparts. 
Surface distribution of X-ray emission is sensitive 
to anisotropy of the explosion and nonuniformity of the interstellar 
medium. 

\medskip \hrule \medskip

}
\vspace{1.0cm}

\baselineskip = 11.2pt

\noindent {\small {\bf INTRODUCTION}}
\medskip

\noindent In some models of supernova explosion the energy is realised 
anisotropically \cite{BK-70,Bodenheimer-Woosly-83}. Aspherical explosion of 
supernova (SN) may cause a barrel-like supernova remnant (SNR) \cite{BK-L-S}. 
Observations show that SN may really be aspherical, as e.g. 
in well-known case of SN1987A 
\cite{Papaliolios-89,Chevalier-Socker-89,Chugaj-91,GMSTRK-97}. 
High birth velocities of pulsars aslo suggest anizotropy in SN explosion 
\cite{Lyne-Lorimer-94}. 
In the past, the morphological evolution and instabilities of 
asymmetric SN explosion were investigated in 
\cite{Blinn,BK-83,Yamada-Sato-91}. 

In present work we consider properties of  
thermal X-ray emission of adiabatic remnants of aspherical SN 
explosions. 
Hydrodynamics is modelled here with an approximate method 
for description of the adiabatic phase of a remnant of aspherical 
SN explosion in an arbitrary large-scale nonuniform medium 
\cite{Hn-Pet-99}. 
Equilibrium thermal X-ray emission model is taken from 
\cite{Raym-Smith77} and approximation for the Gaunt factor from 
\cite{Mewe-Lem-Oord86}. 

\bigskip
\noindent {\small {\bf ASPHERICAL SN EXPLOSION IN A UNIFORM MEDIUM}}
\label{sph_expl-uni_media}
\medskip

\noindent Spherical adiabatic SNR in uniform medium is described by 
self-similar Sedov solutions \cite{Sedov} where  
distributions of thermodynamic parameters are self-similar downstream: 
$\rho(r,t)=\rho_{\rm s}\cdot\overline{\rho}(\overline{r})$,
$P(r,t)=P_{\rm s}\cdot\overline{P}(\overline{r})$, 
$T(r,t)=T_{\rm s}\cdot\overline{T}(\overline{r})$, 
where distance from the center $\overline{r}$, density $\overline{\rho}$, 
pressure $\overline{P}$, temperature $\overline{T}$ are normalized 
on the values at the shock: $R_{\rm s},$ $\rho_{\rm s},$ $P_{\rm s},$ 
$T_{\rm s}.$ 

Let us consider aspherical supernova explosion 
\begin{equation}
\label{E_theta-ord}
E(\vartheta,\phi)=E_o\ \psi(\vartheta,\phi)\quad {\rm where}\quad 
\int\limits_0^{2\pi}d\phi\int\limits_0^\pi
\psi(\vartheta,\phi)sin\vartheta d\vartheta=4\pi. 
\end{equation}
As an example of aspherical energy distribution we will consider the 
function: 
\begin{equation}
\label{E_theta-1}
\psi(\vartheta,\phi)=\left(1-{b\over 2}\right)+b|\cos\vartheta|, \qquad 0\le b< 2.
\end{equation}
SNR 3C~58, which is probably on the free expansion phase of his 
evolution, has maximal axis ratio among known young SNRs: 
$1.67.$ So, possible anisotropy of explosion energy distribution is 
$E_{\rm max}/E_{\rm min}=(R_{\rm max}/R_{\rm min})^2=2.8.$ 
Such anisotropy have place for (\ref{E_theta-1}) if $b\approx 1$. 

Let us consider uniform medium. 
Our hydrodynamic method works within the framework of 
sector approximation approach where flows in sectors 
are independent. This causes essential simplification in modelling 
of the object. Namely, in such a case, we only have  
to re-normalize energy in each sector and 
distributions of parameters in each sectors remain to be self-similar 
since each sector may be described by the Sedov solution with a relevant 
value of explosion energy. 
Thus, the flow characteristics may be written as 
\begin{equation}
\label{asph-prof-ord}
\begin{array}{l}
R_{\rm s}(\vartheta,\phi)=\tilde R_{\rm s}\ \psi(\vartheta,\phi)^{1/5},\qquad
D(\vartheta,\phi)=\tilde D\ \psi(\vartheta,\phi)^{1/5},\\ \\
\rho_{\rm s}(\vartheta,\phi)=\rho_{\rm s},\qquad
P_{\rm s}(\vartheta,\phi)=\tilde P_{\rm s}\ \psi(\vartheta,\phi)^{2/5},\qquad
T_{\rm s}(\vartheta,\phi)=\tilde T_{\rm s}\ \psi(\vartheta,\phi)^{2/5},\\ 
\end{array}
\end{equation}
where $\tilde R_{\rm s},$ $\tilde D,$ $\tilde P_{\rm s},$ 
$\tilde T_{\rm s}$ coinside with relevant values in case of Sedov SNR. 

Luminosity of SNR 
\begin{equation}
\label{L_x-ordinary}
L_{\rm x}=\int\limits_0^{2\pi}d\phi\int\limits_0^{\pi}\sin\vartheta d\vartheta
\int\limits_0^{R_{\rm s}(\vartheta, \phi)}\Lambda(T)n_{\rm e}n_{\rm H}r^2dr
\end{equation}
where emissivity $\Lambda(T)=\int\big(\Lambda_{\rm c}(T,\varepsilon)+
\Lambda_{\rm l}(T,\varepsilon)\big)d\varepsilon$, 
subscript "c" refer to continuum and "l" to line emission. 
It is in case of the Sedov SNR 
\begin{equation}
\label{uni-sph-L}
L_{\rm x}=C(\gamma,\mu)\cdot E_{51}\ n_{\rm H}\ T_{\rm s}^{-1}\cdot 
I_o(T_{\rm s}) \qquad {\rm erg/s},
\end{equation}
where 
\begin{equation}
\label{uni-sph-I_o}
I_o(T_{\rm s})=4\pi\int\limits_0^1
\Lambda\big(T_{\rm s}\cdot\overline{T}(r)\big)\ \overline{n}^2(r)\ r^2\ dr, 
\end{equation}
and $C=3.65\cdot10^{66}$ for $\gamma=5/3,$ $\mu=0.609.$ 
Thermal X-ray spectral index 
\begin{equation}
\label{alpha-def}
\alpha=-{\partial \ln  \over \partial \ln \varepsilon} 
\int\limits_V\Lambda_{\rm c}(T,\epsilon)n_{\rm e}n_{\rm H} dV 
\end{equation}
is, in the case of the Sedov SNR, 
\begin{equation}
\label{uni-sph-a}
\alpha_o=-{\partial\over\partial\ln\varepsilon}\ \ln I_{o,{\rm c}}
(T_{\rm s}),
\end{equation}
where 
\begin{equation}
\label{uni-sph-I_oc}
I_{o,{\rm c}}(T_{\rm s})=\int\limits_0^1
\Lambda_{\rm c}\big(T_{\rm s}\cdot\overline{T}(r)\big)
\ \overline{n}^2(r)\ r^2\ dr.
\end{equation}

\begin{figure}[t]
\epsfxsize=14truecm
\centerline{\epsfbox{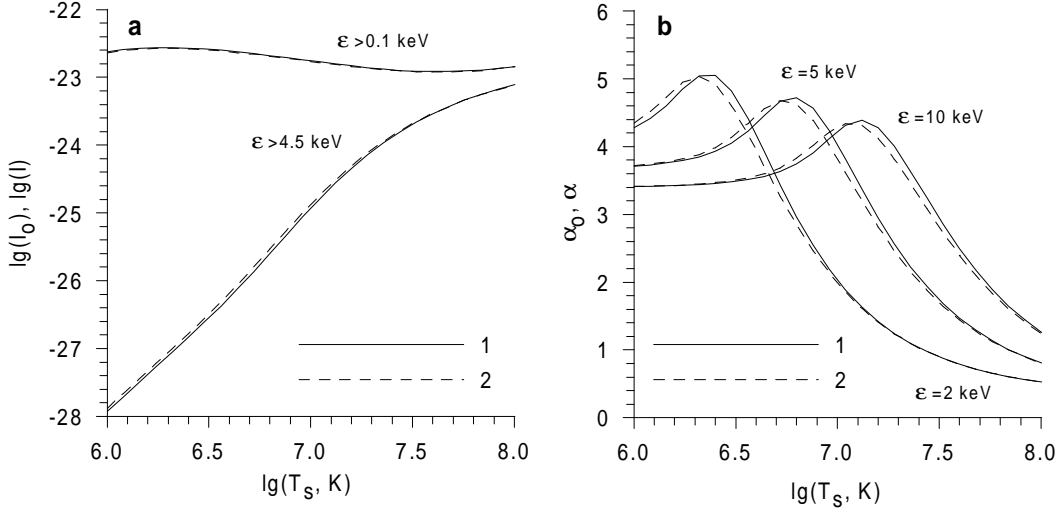}}
\caption[]{{\bf a} Integrals $I_o(T_{\rm s})$ and $I(T_{\rm ch})$ for various 
	emission band and different distribution of supernova explosion energy. 
	1 -- $I_o(T_{\rm s}),$ 
	2 -- $I(T_{\rm ch})$ for $b=1.5.$
	{\bf b} Spectral indexes $\alpha_o(T_{\rm s})$ and $\alpha(T_{\rm ch})$ 
	at different photon energy. 
	1 -- $\alpha_o(T_{\rm s}),$ 
	2 -- $\alpha(T_{\rm ch})$ for $b=1.5.$ 
           }
\label{I_o-I}
\end{figure}

Luminosity of a remnant of aspherical SN explosion 
$E(\vartheta,\phi)=E_o\ \psi(\vartheta,\phi)$ in uniform medium is 
\begin{equation}
\label{uni-asph-L}
L_{\rm x}=C(\gamma,\mu)\cdot E_{51}\ n_{\rm H}\ 
	\tilde T_{\rm s}^{-1}\cdot I(\tilde T_{\rm s}) \qquad {\rm erg/s},
\end{equation}
where integral 
\begin{equation}
\label{uni-asph-I}
I(\tilde T_{\rm s})=\int\limits_0^{2\pi}
d\phi\int\limits_0^{\pi}\psi(\vartheta,\phi)^{3/5}\ \sin\vartheta d\vartheta
\int\limits_0^1\Lambda(\tilde T_{\rm s}\ \psi(\vartheta,\phi)^{2/5}\ 
\overline{T}(r))\ \overline{n}^2(r)\ r^2\ dr.
\end{equation}
Thermal X-ray spectral index of such SNR is 
\begin{equation}
\label{uni-asph-a}
\alpha=-{\partial\over\partial\ln\varepsilon}\ \ln I_c(\tilde T_{\rm s}),
\end{equation}
where   
$I_c(\tilde T_{\rm s})$ is the same integral as (\ref{uni-asph-I}) with 
$\Lambda_c$ instead of $\Lambda$. 
In general case, the temperature 
$\tilde T_{\rm s}$ in 
(\ref{uni-asph-L})-(\ref{uni-asph-a}) 
is a characteristic parameter, 
the value of $\tilde T_{\rm s}$ coincides with a 
real temperature $T_{\rm s}$ on the shock front 
in the case of Sedov SNR.

Since volume, mass and effective temperature of a remnant of aspherical 
SN explosion differ from spherical SNR within $1\%$ only 
\cite{Pet-2000}, 
we may expect that luminosity $L_x$ and spectral index $\alpha$ 
of such SNR will be close to their Sedov counterparts because 
these characteristics essentially depend on emission measure 
$EM\simeq n_{\rm e}^2V\simeq M^2V^{-1}.$ Really, 
Fig.~\ref{I_o-I}a shows integrals $I_o(T_{\rm s})$ and  
$I(\tilde T_{\rm s})$ and Fig.~\ref{I_o-I}b demonstrates variation of 
thermal X-ray spectral indexes $\alpha_o(T_{\rm s})$ and 
$\alpha(\tilde T_{\rm s})$. We see that luminosity in the band 
$\varepsilon>0.1\ {\rm keV}$ has maximal differences of order $2\%$ 
for $b=0.5$ and $5\%$ for $b=1.5;$ they are $2\%$ and $17\%$ in the band 
$\varepsilon>4.5\ {\rm keV}$. Relative differences in spectral index 
are less then $1\%$ for $b=0.5$ and $5\%$ for $b=1.5.$ \par

Thus, summarizing, we cannot distinguish the cases of spherical and 
aspherical supernova explosions in uniform ISM having characteristics 
of SNR as a whole object, because such 
characteristics are very close to their Sedov counterparts. 



It is easy to show that surface brightness along the radius of the 
projection of Sedov SNR distributes as 
\begin{equation}
\label{S_uni_sph}
S_o(x)={\rm const}\cdot E_{51}^{1/3}n_{\rm H}^{5/3}T_{\rm s}^{-1/3}
\int\limits_x^1\Lambda(T_{\rm s}\overline{T}(r))\ 
{\overline{n}^2(r)\ dr\over \sqrt{r^2-x^2}}
\end{equation}
where $x$ is the position along the SNR radius of the projection in the 
units of the radius ($0\leq x\leq 1$), ${\rm const}=1.1\cdot 10^{21}$
for $\gamma=5/3$ and $\mu=0.609$. 

If explosion is axially-symmetrical $E(\vartheta,\phi)=E_o\psi(\vartheta)$ 
($\partial E/\partial\phi=0$) 
then the shape of the shock front is a figure of revolution with profile 
$R(\vartheta)=\tilde R_{\rm s}\psi(\vartheta)^{1/5}.$ 
If, additionally, the symmetry axis of the distribution 
$\psi(\vartheta)$ lies in the plan of projection 
(inclination angle $\delta=0$), the  
surface brightness distribution of such SNR is given with 
\begin{equation}
\label{S_uni_asph_1}
S(x,\ z)=S_o\psi(\vartheta)^{1/5}\sin\vartheta,
\end{equation}
where $z=R(\vartheta)\cos\vartheta$ is the coordinate along the symmetry 
axis of the distribution $\psi(\vartheta)$, 
$x$ is coordinate perpendicular to $z$ in the plane of projection. 
This indicates that the surface brightness 
distribution profiles parallel to axis $x$ are re-normalzed profiles 
of the surface brightness distribution in Sedov SNR. 
This fact may be used to test the orientation of
axis-symmetrical SNRs in uniform medium. 
Distribution of the spectral index has a similar behevior in case of 
$\delta=0^o$: 
profiles parallel to axis $x$ are like to profile of the Sedov SNR. 
Cases $\delta\neq 0^o$ are considered in \cite{Pet-2000}. 

Evolution of an adiabatic remnant of aspherical SN explosion in 
nonuniform medium is investigated in a further paper \cite{Pet-2000}. 

\bigskip
\noindent {\small {\bf CONCLUSIONS}}
\medskip

\noindent 
Aspherical supernova explosion causes only small differences 
in thermal X-ray limonosity and spectral index of adiabatic SNR 
comparing to a spherical explosion case. Such a behaviour takes 
place in different X-ray bands and for a spectral index at different 
frequencies in X-rays. Thus, the thermal X-ray spectrum of a remnant 
of aspherical explosion is close to the spectrum of Sedov SNR. 

These facts do not allow us to distinguish between the cases of 
spherical and aspherical SN explosion if we consider only the 
integral characteristics of X-ray emission of an adiabatic SNR. 

X-ray properties of an SNR essentially depend on features of a 
disturbed plasma flow behind the shock because 
emission depends on density squared and temperature on  
velocity squared. Therefore, 
surface distributions of X-ray emission charateristics are a 
sensitive test on the conditions in which the object evolves. 
Asphrericity of explosion reveals itself in surface distributions of 
thermal X-ray characteristics. 
In the case of an axially-symmetrical explosion in uniform medium 
and when the axis of symmetry 
lies in the plan of projection, the profiles of X-ray surface 
brightness and spectral index along the lines perpendicular to the axis 
are re-normalized profiles of relevant distributions of the Sedov SNR. 

\bigskip


\end{document}